\documentclass[]{spie}  %>>> use for US letter paper
\usepackage[]{graphicx,journals}

\title{The interferometric baselines and GRAVITY astrometric error budget} 

\author{S. Lacour\supit{a}, F. Eisenhauer\supit{b}, S. Gillessen\supit{b}, O. Pfuhl\supit{b}, Y. Kok\supit{b}, G. Perrin\supit{a}, K. Rousselet-Perraut\supit{c}, C. Straubmeier\supit{d}, W. Brandner\supit{e}, A. Amorim\supit{f}, J. Woillez\supit{g}
 and H. Bonnet\supit{g}
\skiplinehalf
\supit{a} LESIA/Observatoire de Paris, 5 place Jules Janssen, 92195 Meudon,
France \\
\supit{b} Max-Planck Institut fur extraterrestrische Physik (MPE), 85748, Garching, Germany \\
\supit{c} IPAG, Univ. Grenoble Alpes, CNRS, F-38000 Grenoble, France  \\
\supit{d} Physikalisches Institut, Universitat zu Koln, Zulpicher Str. 77, 50937 Koln, Germany\\
\supit{e} Max-Planck-Institut fur Astronomie, Konigstuhl 17, 69117 Heidelberg, Germany \\
\supit{f} SIM and FCUL - Edificio C8, Campo Grande, P-1749-016 Lisboa, Portugal \\
\supit{g} European Southern Observatory, Karl-Schwarzschild-Str. 2, 85748, Garching, Germany\\
}

\authorinfo{Email: sylvestre.lacour@obspm.fr}
\begin{document} 
  \maketitle 

%%%%%%%%%%%%%%%%%%%%%%%%%%%%%%%%%%%%%%%%%%%%%%%%%%%%%%%%%%%%% 
\begin{abstract}
GRAVITY is a new generation beam combination instrument for the VLTI. Its goal is to achieve microarsecond astrometric accuracy between objects separated by a few arcsec. This $10^6$ accuracy on astrometric measurements is the most important challenge of the instrument, and careful error budget have been paramount during the technical design of the instrument. In this poster, we will focus on baselines induced errors, which is part of a larger error budget.
\end{abstract}

%>>>> Include a list of keywords after the abstract 


%%%%%%%%%%%%%%%%%%%%%%%%%%%%%%%%%%%%%%%%%%%%%%%%%%%%%%%%%%%%%
\section{INTRODUCTION}
\label{sec:intro}  % \label{} allows reference to this section

The GAIA mission\cite{2001A&A...369..339P} is launched and, at the time of writing, is awaiting its first results. It is expected to reach accuracies of several micro arcseconds on bright stars. However, a long baseline interferometer still have relevance. It can do differential astrometry in crowed field, or between two close objects of large magnitude difference. This is necessary to do astrometry on the galactic center for exemple. Or to give precise astrometry on the orbit of exoplanets next to their host stars.

The goal of the GRAVITY\cite{2011Msngr.143...16E} instrument is to reach accuracies of 10\,$\mu$as between two targets of magnitudes 10 and 16, separated by one or two arcseconds. We are therefore talking about differential astrometry. It will combine the 4 8-meter VLT telescopes at Paranal, benefiting from interferometric baselines of the order of 100 meters. The interferometer will observe simultaneously the two targets ($\vec s$ and $\vec p$), and measure the optical path difference (OPD) between the two interferometric arms and target to derive the astrometry. It is therefore a double difference optical path length measurement. In its simplest form, the OPD is related to the astrometric position of the stars by the relation:
\begin{equation}
 \delta OPD_{\rm simple}=(\vec s - \vec p ) \cdot \vec B\,,
\end{equation}
where $\vec B$ is the baseline vector.

However, one of the difficulty of long baseline interferometry for astrometry is the knowledge of the baseline vector. To reach 10\,$\mu$as accuracy between two targets separated by 1\,arcsecond, the baseline length has to be known within a few parts per million (ppm). It means knowing the 100\,m baselines of the Very Large Telescope Interferometer (VLTI) to a sub-millimeter level. To do so, we first have to define precisely what is the baseline, and exactly establish what physical quantity determine its length. This is the goal of section~\ref{sec:base}.

\section{THE BASELINES}
\label{sec:base}

   \begin{figure}
   \begin{center}
   \begin{tabular}{c}
   \includegraphics[height=7cm]{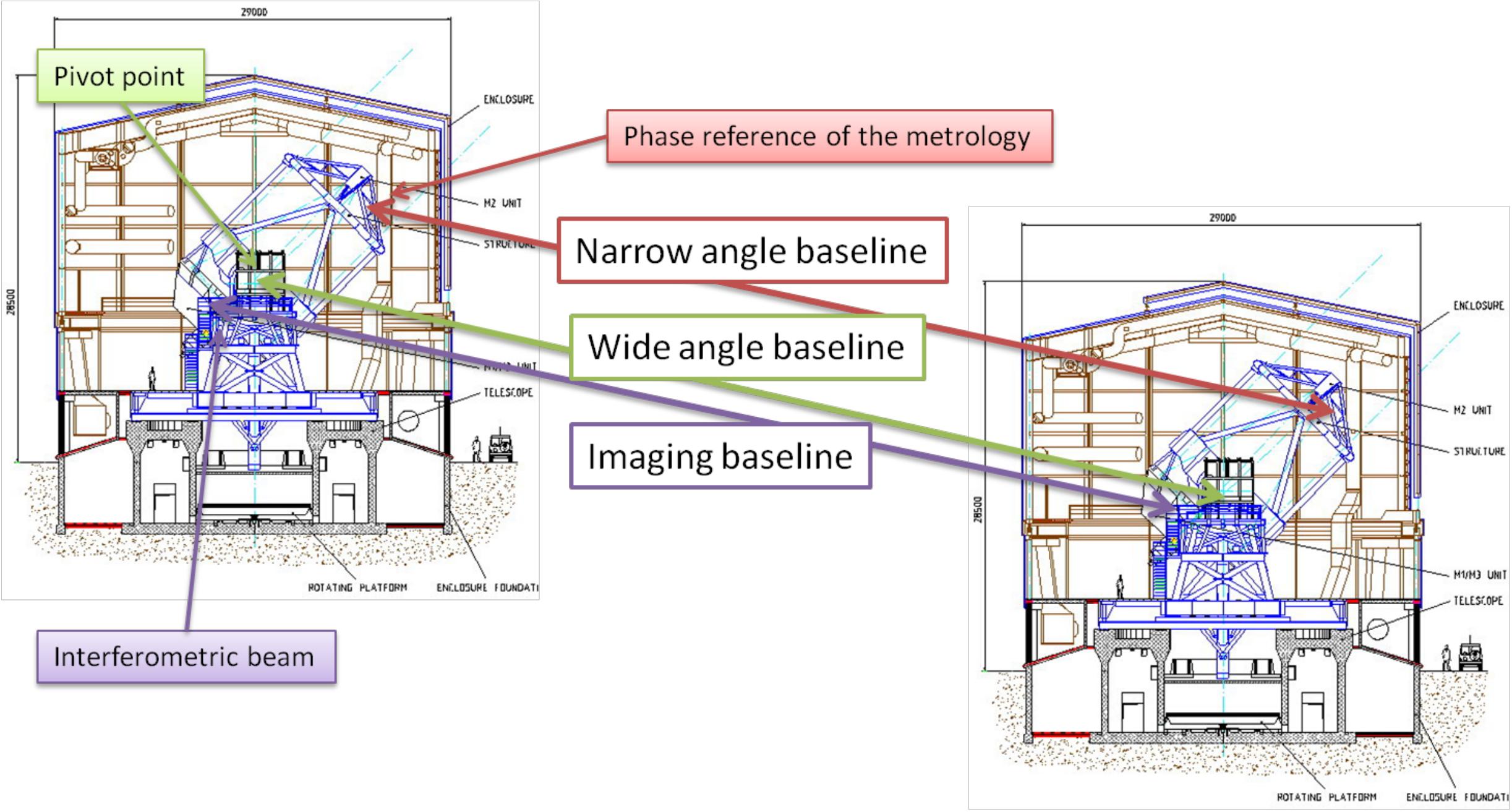}
   \end{tabular}
   \end{center}
   \caption[example] 
%>>>> use \label inside caption to get Fig. number with \ref{}
   { \label{fig:baselines} 
The three baselines defined by Woillez \& Lacour (2014)\cite{PI}. The baselines are vectors, but are not defined by their length and orientation, but by their limit points in geographical space.
}
   \end{figure} 
%-------------

We can distinguish three baselines in an interferometer\cite{PI}. The baselines are defined by their limit points. The baselines vector of the three baselines can be identical, but not necessarily. The three baselines are:
\begin{itemize}
\item The wide angle baseline (WAB), the baseline which is delimited by the position of the telescope pivot point. The WAB is used for wide angle astrometry.
\item The narrow angle baseline (NAB), the baseline which is delimited by the metrology end points. If several end points are used for the metrology, and the OPD is averaged between all of them, the end point of the metrology is the geographical average of each of them. The NAB is used for narrow angle astrometry (as it is the case for GRAVITY).
\item The imaging baseline (IMB), the baseline which is delimited by the position of the interferometric pupil in each telescope. The interferometric pupil is the pupil of the beam combiner pupil (the one in the lab) weighted by the pupil of the telescope. The pupil position is situated in three dimensions. For exemple, in geographical coordinate the pupil is often several hundred of meters below the ground.
\end{itemize}
The three baselines are represented in figure~\ref{fig:baselines}.

Each one of the baseline limit points has properties which are useful, and which make these 3  baseline definitions useful:
\begin{itemize}
\item The telescope pivot point is the center of rotation of the telescope. It does not move when the telescope change target\footnote{except in the case of telescope flexure}. Its position does not depend on telescope-pointing nor the optical path inside the interferometer.
\item The metrology end points are dependent on the way the metrology is build. In the case of gravity, the metrology end-point is situated on the spider arms, and therefore moves with the telescope (it is fixed in the telescope reference frame).
\item The imaging baseline limit points are the pupil of the interferometer (they can be different from the telescope pupil). The important property of these limit points are the fact that they are field invariant: they do not move when the path of the starlight inside the interferometer change. Similarly, the OPD seen by the metrology up to the pupil does not depend on its optical path (or beam walk).
\end{itemize}
It is these properties which will be used below to establish an analytical expression of the astrometric measurement as a function of the 3 dimensional positions of the limit points.

\section{ANALYTICAL EXPRESSION OF THE OPD MEASUREMENT}

   \begin{figure}
   \begin{center}
   \begin{tabular}{c}
   \includegraphics[height=7cm]{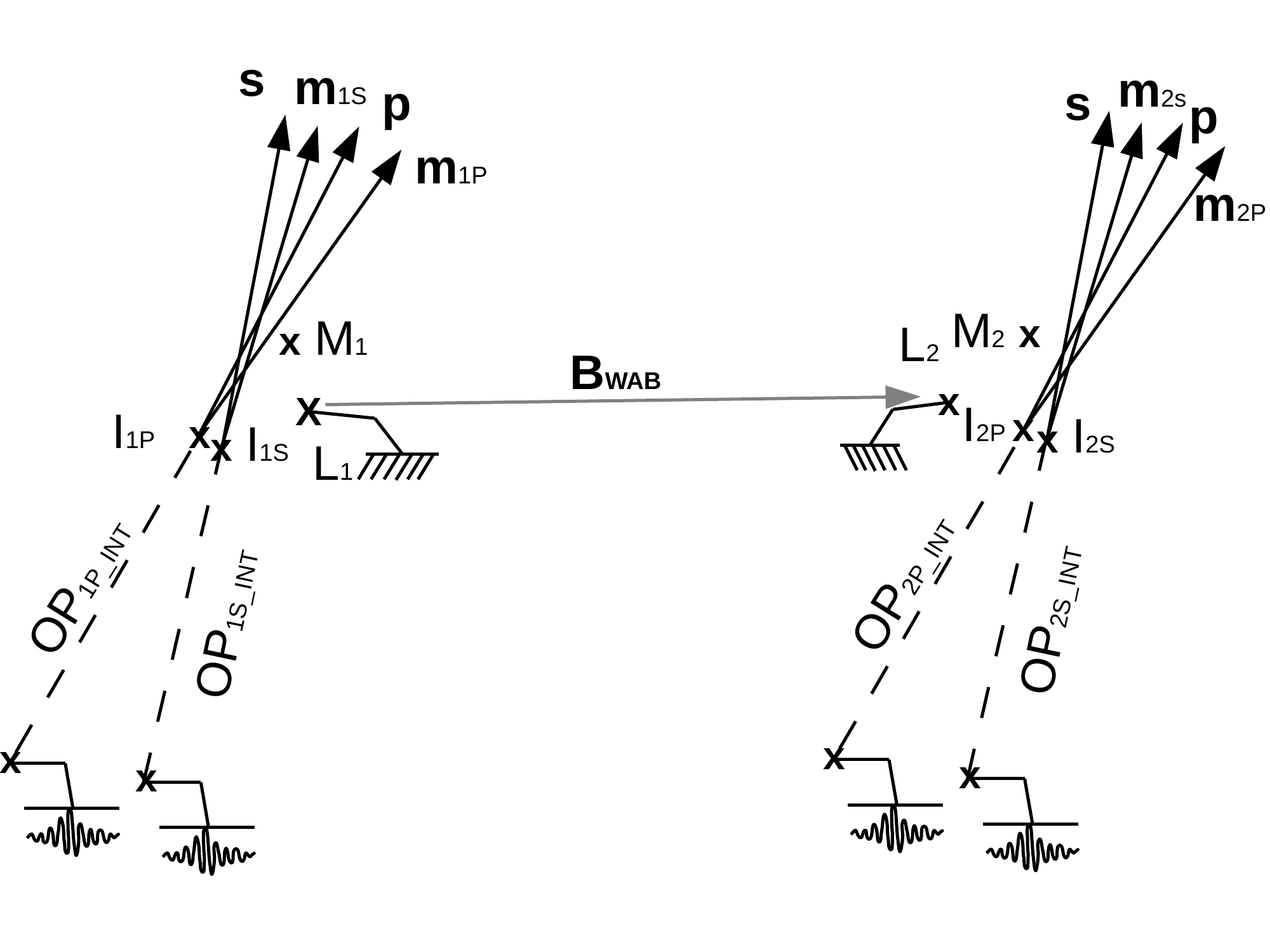}
   \end{tabular}
   \end{center}
   \caption[example] 
%>>>> use \label inside caption to get Fig. number with \ref{}
   { \label{fig:interferometer} 
Representation of a 2 telescopes interferometer without any mirror. The optical paths are virtual. Each one of the two beam combiners (for target $\vec s$ and $\vec p$) has two virtual representation in this diagram: one through each telescope.}
   \end{figure} 

A simplified representation of a 2-telescope interferometer is represented in figure~\ref{fig:interferometer}. To simplify the interferometer to its minimum, all the mirrors have been removed. The representation therefore represent the virtual rays as straight lines, from the stars down to the beam combiners. There are two beam combiners: one for star $\vec s$ and one for star $\vec p$. Each one of these two beam combiners is represented twice in figure~\ref{fig:interferometer}. This is due to the fact that the star "sees" the beam combiner twice, inside each telescope aperture. 

The limit points of all the baselines discussed in section~\ref{sec:base} are represented in this figure. The pivot points of the telescopes are shown as $L_1$ and $L_2$. The end points of the metrology are represented as $M_1$ and $M_2$. Last, the instrument pupils are represented as $I_{1P}$, $I_{2P}$, $I_{1S}$ and $I_{2S}$. Note that there are two imaging baselines, and four imaging baseline limit points. This is because the two pupils of beam combiner 1 are not necessarily the same as the pupils of beam combiner 2.

To derive the analytical expression of the metrology measurements, three sets of equations are necessary:
\begin{enumerate}
\item The optical path length from the star to the white light fringe is equal between each arm of the interferometer:
\begin{eqnarray}
 OP_{1S-EXT} + OP_{1S-INT} &= &  OP_{2S-EXT} + OP_{2S-INT} \\
 OP_{1P-EXT} + OP_{1P-INT} &= &  OP_{2P-EXT} + OP_{2P-INT}
\end{eqnarray}
\item The external optical path length difference is equal to the scalar product between the targets and the imaging baselines:
\begin{eqnarray}
 OP_{1S-EXT} - OP_{2S-EXT} = \vec  s \cdot \overrightarrow{ { I_{1S}I_{2S} }} \label{OPDMM1}& \\
 OP_{1P-EXT} - OP_{2P-EXT} = \vec  p \cdot \overrightarrow{ { I_{1P}I_{2P}  }} \label{OPDMM2}&.
\end{eqnarray}
\item The metrology measurement $\delta OPD_M$ is a double difference: a difference between the optical length observed between each arms of the interferometer,  and between the two targets $\vec s$ and $\vec p$. The measurement is equal to the internal optical path plus a length which correspond to the distance between the pupil and the metrology end points. This additional optical length depends on the direction of propagation of the metrology $\vec m$ in each arm. It is equal to $\vec { m_{1S}} \cdot  \overrightarrow{  { I_{1S}M_1}}$ for telescope $1$ and star $\vec s$. The total metrology measurement is therefore:
\begin{eqnarray}
\delta OPD_{M} &=& OP_{2S-INT}  +  \vec { m_{2S}} \cdot  \overrightarrow{  { I_{2S}M_2}} \nonumber  \\
&&- OP_{1S-INT} -  \vec { m_{1S}} \cdot \overrightarrow{ { I_{1S}M_1}} \nonumber  \\
&&+  OP_{1P-INT} +  \vec { m_{1P}} \cdot \overrightarrow{  { I_{1P}M_1}} \nonumber  \\
&&- OP_{2P-INT} -  \vec { m_{2P}} \cdot  \overrightarrow{ { I_{2P}M_2}} \, .  \label{OPDII0}
\end{eqnarray}
\end{enumerate}

By combining these three sets of equations, we can derive the expression of the OPD measured by the metrology as a function of the stars $\vec s$ and $\vec p$:
\begin{eqnarray}
\delta  OPD_{M} &=& (\vec s - \vec p) \cdot (\overrightarrow{ {M_1L_1}} +  \overrightarrow { B_{\rm WAB}} + \overrightarrow{ {L_2M_2}} ) \nonumber \\
 &&+ (\vec { m_{2S}} - \vec  s) \cdot \overrightarrow{ {I_{2S}M_2}} -  (\vec {m_{1S}} - \vec s) \cdot \overrightarrow{ {I_{1S}M_1}}    \nonumber \\
 && +  (\vec { m_{1P}} - \vec p) \cdot \overrightarrow{ {I_{1P}M_1}} -  (\vec {m_{2P}} - \vec p) \cdot \overrightarrow{ { I_{2P}M_2}} \,,
 \label{OPDIMG}
\end{eqnarray}
where $\overrightarrow{ {M_1L_1}} +  \overrightarrow { B_{\rm WAB}} + \overrightarrow{ {L_2M_2}}=\overrightarrow{ {M_1M_2}} = \overrightarrow { B_{\rm NAB}}$ is the narrow angle baseline. The lasts four terms are second order terms due to difference in beam walk between the metrology and star light, combined with limit points mismatches. 

\section{ERROR BUDGET}

\subsection{Errors Related to the Determination of the Narrow Angle Baseline}

The narrow angle baseline is:
\begin{equation}
\overrightarrow { B_{\rm NAB}} = \overrightarrow{ {M_1L_1}} +  \overrightarrow { B_{\rm WAB}} + \overrightarrow{ {L_2M_2}}\,.
\end{equation}
The main difficulty to establish a precise narrow angle baseline is to be able to accurately switch reference frames:
\begin{itemize}
\item The metrology end points are set in the telescope reference frame (on the spider arms)
\item The pivot points are known in the terrestrial reference frame (on the ground)
\item But the narrow angle baseline has to be know in the sidereal reference frame
\end{itemize}

Changing the $\overrightarrow{ {M_1L_1}}$ and $\overrightarrow{ {L_2M_1}}$ vectors from telescope to sidereal reference frame requires to know the physical pointing of the telescope. Modern telescopes can point with an accuracy below $10"$, meaning an error of  $10"$ between the physical pointing and the optical pointing. Since in the GRAVITY case the metrology sensors are around 8 meters above the pivot points ($|\overrightarrow{LM}|=8\,$m), the resulting baseline error is $10" \times 8\,$m$=0.5\,$mm.

Changing $\overrightarrow { B_{\rm WAB}}$  from terrestrial to sidereal reference frame requires precise timing of the observations, and precise knowledge of the exact sidereal time at the position of the observatory. For example, moving 100\,m west at the Paranal observatory delay the sidereal time by 1/4 of a second. We assume that we we will be able to time our observations with respect to the sidereal time with an accuracy of 1/15 of a seconds. For $|\overrightarrow{B_{\rm WAB}}|=100\,$m, the resulting error is $1/(15*24*3600)\times 100\,$m$=0.5\,$mm.

Finally, the structural stability of the WAB matters. We estimate it to be also of the order of 0.5\,mm (except during earthquakes).

With respect to a 100\,m baseline, a 0.5\,mm error in baseline will correspond to an error in the astrometric accuracy of 5\,ppm, hence 5$\,\mu$as for two stars separated by 1". The NAB indetermination factors are summarized in the upper part of Table 1.

\subsection{Errors Due to Metrology Beam Walk Combined with Pupil Offsets}

\begin{table*}
\footnotesize
\caption{GRAVITY astrometric error budget (only the part related to the baseline errors)}        
\label{table:1}     
\centering                      
\begin{tabular}{l c c c c}     
\hline\hline            
Error label & Cause & Uncertainties & Consequence & $\mu$as \\  
\hline                     
&\bf $\overrightarrow { B_{\rm NAB}}$ Error  \\
$\overrightarrow { LM}$ in sidereal reference frame & Telescope physical pointing & $\pm 10"$ & Baseline error 0.5\,mm & 5 \\
$\overrightarrow { B_{\rm WAB}}$ in sidereal reference frame & LST time during observation & 1/15\,s  & Baseline error 0.5\,mm & 5\\
$\overrightarrow { B_{\rm WAB}}$ stability  & Paranal structural stability & 0.5\,mm  & Baseline error 0.5\,mm & 5\\
\hline          
&\bf Pupil Error\\
Lateral pupil  & $ (\vec  m - \vec  s) \cdot (\overrightarrow{\vec  {IM}} \perp z)$ &  10mas -- 4cm & OPD error of 2nm & 4 $\sqrt 4$ \\
Longitudinal pupil  & $ (\vec m - \vec  s) \cdot (\overrightarrow{\vec  {IM}} \parallel z)$ &  10mas  -- 10km & OPD error of 10pm & 0.02$\sqrt 4$  \\
\hline
\end{tabular}
\footnotesize {Accuracy obtained between two targets separated by 1 arcsec, assuming a single 100\ m baseline
}
\end{table*}

It requires both a metrology beam walk error ($\vec m_S-\vec s\neq\vec 0$ or $\vec m_P-\vec p\neq\vec 0$) with a pupil error ($\overrightarrow{{IM}}\neq\vec 0$) to create this additional error term. To keep the scalar product of these two terms under control, the GRAVITY design team has made two technical choices:
\begin{enumerate}
\item The metrology beam\cite{2012SPIE.8445E..1OG} is launched by the same single mode fibers that receive the stellar light. This way, since the position of the fiber is aligned with the stellar beam to maximize the injection, the metrology is also aligned to the stellar beam. The beam walk will be the same, to the precision obtained when optimizing the coupling. When using the UTs, the on-sky single mode core is around 60\,mas. We assume that after optimization, the error in the propagation direction between the two beams will be below 10\,mas.
\item To limit the offsets between the interferometric pupil position and the metrology end point, the acquisition camera\cite{2012SPIE.8445E..34A} inside GRAVITY cryostat constantly track 4 beacons mounted on the spider arms, next to the metrology sensors. This acquisition camera can track the pupil with a precision of 4\,cm laterally, and 10\,km longitudinally\footnote{10\,km in the 8\,m beam is obtained by tracking longitudinally 1\,m in the 80\,mm beam situated at the entrance of the cryostat}.
\end{enumerate}

It is noteworthy to remark that both the metrology end points ($M$) and the pupil reference points ($I$) are situated in a 3 dimensional space, and that the longitudinal error between the two can be extremely large (10\,km). The error terms due to pupil error and metrology beam walk can therefore be separated into two terms: 
 \begin{equation}
 (\vec  m - \vec  s) \cdot (\overrightarrow{\vec  {IM}} \perp z) = \sin (10\,{\rm mas}) \times 4\,{\rm cm} = 2\,{\rm mm}
 \end{equation}
  for the lateral pupil shift, and 
 \begin{equation}
 (\vec  m - \vec  s) \cdot (\overrightarrow{\vec  {IM}} \parallel z) =  (1- \cos (10\,{\rm mas}) )\times 10\,{\rm km} = 10\,{\rm pm}
 \end{equation}
  for the longitudinal pupil shift.
  
 Translated into an astrometric error, and assuming a projected baseline of 100\,m, these two OPD errors are equivalent to respectively 4 and 0.02\,$\mu$as. These values have to be multiplied by $\sqrt 4$ to account for the presence of 4 of such terms in Equation~(\ref{OPDIMG}). 

\section{CONCLUSION}

Table~\ref{table:1} summarize the different errors which are caused by baselines errors. The first part is caused by instabilities of the NAB.  
The second part is due to a complex cross-talk between the imaging baseline and the narrow angle baseline. It includes analysis of the impact of the 3 dimensional nature of the position of the interferometric pupil.
A more extended error budgets, as well as more details, can be found in the paper by Lacour et al.\,(2014)\cite{2014arXiv1404.1014L}.

%%%%%%%%%%%%%%%%%%%%%%%%%%%%%%%%%%%%%%%%%%%%%%%%%%%%%%%%%%%%%
%\acknowledgments     %>>>> equivalent to \section*{ACKNOWLEDGMENTS}       
 
%This unnumbered section is used to identify those who have aided the authors in understanding or accomplishing the work presented and to acknowledge sources of funding.  

%%%%%%%%%%%%%%%%%%%%%%%%%%%%%%%%%%%%%%%%%%%%%%%%%%%%%%%%%%%%%
%%%%% References %%%%%

\bibliography{report,GRbib}   %>>>> bibliography data in report.bib
\bibliographystyle{spiebib}   %>>>> makes bibtex use spiebib.bst

\end{document}